%% file: manuscript.tex
\documentclass[natbib=true, screen, sigconf]{acmart}
\graphicspath{ {./images/} }

\AtBeginDocument{%
  \providecommand\BibTeX{{%
    \normalfont B\kern-.5em{\scshape i\kern-.25em b}\kern-.8em\TeX}}}

\usepackage{enumitem}
\setlist[itemize]{leftmargin=*}
\usepackage{graphicx}
\usepackage{caption}

\usepackage{subcaption}

\usepackage{placeins}
\usepackage{float}
\usepackage{multirow}
\usepackage{balance}
\usepackage{xspace}
\usepackage{utfsym}
\usepackage{pifont}

\input{macros.tex}

\copyrightyear{2024}
\acmYear{2024}
\setcopyright{rightsretained}
\acmConference[SIGIR '24]{Proceedings of the 47th International ACM SIGIR Conference on Research and Development in Information Retrieval}{July 14--18, 2024}{Washington, DC, USA}
\acmBooktitle{Proceedings of the 47th International ACM SIGIR Conference on Research and Development in Information Retrieval (SIGIR '24), July 14--18, 2024, Washington, DC, USA}
\acmDOI{10.1145/3626772.3657793}
\acmISBN{979-8-4007-0431-4/24/07}
\makeatletter
\gdef\@copyrightpermission{
   \begin{minipage}{0.3\columnwidth}
     \href{https://creativecommons.org/licenses/by-nd/4.0/}{\includegraphics[width=0.90\textwidth]{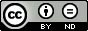}}
   \end{minipage}\hfill
   \begin{minipage}{0.7\columnwidth}
     \href{https://creativecommons.org/licenses/by-nd/4.0/}{This work is licensed under a Creative Commons Attribution-NoDerivs International 4.0 License.}
   \end{minipage}
   \vspace{5pt}
}
\makeatother

\begin{document}

\title[Characterizing Information Seeking Processes with Multiple Physiological Signals]{Characterizing Information Seeking Processes with\\Multiple Physiological Signals}

\author{Kaixin Ji}
\orcid{0000-0002-4679-4526}
\affiliation{%
  \institution{RMIT University} 
  \city{Melbourne}
  \country{Australia}
}
\email{kaixin.ji@student.rmit.edu.au}

\author{Danula Hettiachchi}
\orcid{0000-0003-3875-5727}
\affiliation{
\institution{RMIT University}
  \city{Melbourne}
  \country{Australia}
}

\email{danula.hettiachchi@rmit.edu.au}

\author{Flora D.~Salim}
\orcid{0000-0002-1237-1664}
\affiliation{%
  \institution{The University of New South Wales} 
  \city{Sydney}
  \country{Australia}
}
\email{flora.salim@unsw.edu.au}

\author{Falk Scholer}
\orcid{0000-0001-9094-0810}
\affiliation{
\institution{RMIT University}
  \city{Melbourne}
  \country{Australia}
}
\email{falk.scholer@rmit.edu.au}

\author{Damiano Spina}
\orcid{0000-0001-9913-433X}
\affiliation{
\institution{RMIT University}
  \city{Melbourne}
  \country{Australia}
}
\email{damiano.spina@rmit.edu.au}

\setlength{\intextsep}{10pt plus 2pt minus 2pt}

\begin{abstract}
Information access systems are getting complex, and our understanding of user behavior during information seeking processes is mainly drawn from qualitative methods, such as observational studies or surveys.
Leveraging the advances in sensing technologies, our study aims to characterize user behaviors with physiological signals, particularly in relation to cognitive load, affective arousal, and valence. We conduct a controlled lab study with 26 participants, and collect data including Electrodermal Activities, Photoplethysmogram, Electroencephalogram, and Pupillary Responses. 
This study examines informational search with four stages: the realization of Information Need (\ifn), Query Formulation (\qf), Query Submission (\qs), and Relevance Judgment (\rj). We also include different interaction modalities to represent modern systems, e.g., \qs by text-typing or verbalizing, and \rj with text or audio information.
We analyze the physiological signals across these stages and report outcomes of pairwise non-parametric repeated-measure statistical tests. The results show that participants experience significantly higher cognitive loads at \ifn with a subtle increase in alertness, while \qf requires higher attention. \qs involves demanding cognitive loads than \qf. Affective responses are more pronounced at \rj than \qs or \ifn, suggesting greater interest and engagement as knowledge gaps are resolved. 
To the best of our knowledge, this is the first study that explores user behaviors in a search process employing a more nuanced quantitative analysis of physiological signals.
Our findings offer valuable insights into user behavior and emotional responses in information seeking processes. We believe our proposed methodology can inform the characterization of more complex processes, such as conversational information seeking.

\end{abstract}

\settopmatter{printfolios=true}


\maketitle

\section{Introduction}

One of the core concepts studied in Interactive information retrieval (IIR) is the continuous \cite{ruthven2011interactive}, problem-solving \cite{belkin1980anomalous, kuhlthau2005information, cole2011theory} process around information. 
Over the decades, theoretical models~\cite{kuhlthau2005information, nahl2007social, belkin1980anomalous, taylor1968question, saracevic1997studyingI, marchionini1995information} have attempted to characterize the interactions between users (searchers) and (search) systems from different perspectives.
As outlined by \citet{cole2011theory}, the common search system is ``command-based''~\cite{taylor1968question}, which assumes that users already know what they are looking for (``known answers'') and provide specific requests (``commands'') accordingly, rather than descriptive questions with ``unknown answers''. To come up with this search request, \citeauthor{taylor1968question} 
theorizes that information seeking is a process that transfers from the latter to the former \cite{taylor1968question}; in other words, digging deeper to uncover a more visceral level of need.
This is similar to \citeauthor{kuhlthau2005information}'s proposition that the cognitive state shifts from vague and ambiguous to clear and focused \cite{kuhlthau2005information}.
But both models convey the process as an interchange between affective and cognitive states, driven by a \emph{feeling of uncertainty} and subsequent reactions with physical actions. 
Likewise, \citet{nahl2007social} narrates the exchanges among affect, cognition, and physical actions but emphasizes the role of appraisal as the drive.
Overall, a search begins when users realize their inability or insufficiency of knowledge to solve a problem, prompting them to use a search engine. Each search session may contain multiple iterations of entering and executing queries, assessing search results, and evaluating the information quality. If users are unsatisfied with the collected information, they may reformulate the query and start another iteration \cite{marchionini1995information, kuhlthau2005information, sutcliffe1998towards}.

Theoretical models have traditionally been formulated based on qualitative methods, such as observational studies and surveys, or facial expression analysis (e.g., \cite{kuhlthau2005information, arapakis2009using, lopatovska2014toward, mcduff2021affective}). 
By examining behavioral data and self-ratings, \citet{gwizdka2010distribution} reports that the distribution of mental demand (\emph{cognitive load}) varies across different search stages. 
These observational approaches have limited ability to capture the real situation at a detailed level \cite{lopatovska2014toward}. Some affective activities happen but are not strong enough to be perceived by humans \cite{savolainen2015interplay}. 
This might cause most experiments that rely on observations to find neutral affect as the most frequent during search interaction \cite{lopatovska2014toward, mcduff2021affective}. 
The advancement of physiological sensors presents an opportunity to revisit and refine existing theoretical models \cite{lopatovska2011theories}. 
In information searching or browsing, wearable sensors have been employed to detect user's interests~\cite{White2017881}, satisfaction~\cite{wu2017predicting}, and engagement~\cite{ashlee2016engaged}. It has also been shown that sensor data can indicate affective appraisal (i.e., the continuous interplay between emotions and body perception of surroundings \cite{savolainen2015interplay, daley2014emotional}) in reading comprehension, for example, inferring a sense of preparedness, confidence, and activation of background knowledge when beginning reading \cite{daley2014emotional}.

This paper aims to validate and summarize human factors in theoretical information-seeking models and existing findings. We revisit some of the phenomena observed in the literature by considering the use of physiological data. The physiological data are captured by wearable sensors, including Electrodermal Activity (EDA), Photoplethysmogram (PPG), Electroencephalogram (EEG), and Pupil Dilation (PD). 
Due to the complex nature of information activities and the sensitivity of physiological sensors \cite{ji2023examining}, we conduct a highly controlled lab study to eliminate confounding variables as much as possible. We carefully scrutinize the study materials to minimize the influences of attitudes (relating to cognitive bias) and relevance. Our experimental design is inspired by the experiment by \citet{moshfeghi2018search}. The novel hypotheses that we formalize and explore in this work are built upon the synthesis of established theoretical models and existing empirical results. 
This study focuses on four search stages in a single iteration: the realization of Information Need (\ifn), Query Formulation (\qf), Query Submission (\qs), and Relevance Judgment (\rj).
Further, to account for diverse text- and voice-based systems, we include study conditions around different modalities of presenting and receiving information. In particular, a system receives queries or presents information in text or audio. 
Although \qf and \qs are usually consecutive stages in real-world scenarios, the literature suggests that their underlying activities diverge (discussed later in Section~\ref{sec:hypotheses}), especially when considering the impact of interaction modalities \cite{ji2023towards}. Hence, we treat them as separate stages in this study.

Overall, our results show that \ifn encounters higher cognitive loads and alertness, suggesting the update of knowledge gaps, than \qf. And \qf requires less cognitive demand but enhanced affective feelings than \qs.
Our study also observes more pronounced affective feelings at \rj. This reaction may be linked to the resolution of knowledge gaps, leading to increased interest and engagement. This study complements the understanding of cognitive activities and affective responses during information seeking by offering a detailed perspective with physiological signals.
To the best of our knowledge, this is the first study in IIR to collect and analyze multi-modal physiological data during interactive information search.
The main contributions of our work are three-fold:
\begin{itemize}[topsep=0pt]
    \item Through a comprehensive analysis of literature in the areas of IIR, cognitive science, and affective and wearable computing, we formalize a novel set of hypotheses that allow us to study how search stages can be characterized with physiological signals.
    \item Our proposed controlled lab study design, allowed us to validate (either fully or partially) some of the hypotheses, while also obtaining insights into the rejected ones. This complements our existing knowledge of the role of cognitive and affective activities during the search stages of an information seeking process.
    \item Our study fills the gap of employing physiological wearable sensors in IIR. It can serve as a groundwork for future experiments using physiological sensors to characterize more complex search processes such as conversational information seeking with Large Language Model-based systems.
\end{itemize}

\section{Literature Review \& Hypotheses}
\label{sec:hypotheses}

Information-seeking models have been extensively studied in the field of information retrieval \cite{belkin1980anomalous,marchionini1995information,kuhlthau2005information}. Although some work has aimed to understand the different search stages from a cognitive and affective point of view \cite{gwizdka2010distribution, moshfeghi2018search, mcduff2021affective, Gwizdka2017temporal, arapakis2008affective}, little work has been done to characterize search processes with physiological signals captured from wearable devices \cite{shovon2015search, arapakis2009using}. In this section, we draw attention to theories and findings that exist at the intersection of interactive information retrieval, cognitive science, and affective and wearable computing. Following the recommendation by \citet{riedl2014towards} to assure methodological rigor, we identify the hypotheses in terms of three low-level physiological constructs: \emph{cognitive load}, \emph{affective arousal}, and \emph{affective valence}, and aim to validate them using the quantifiable physiological signals.

We start by summarizing how search stages are conceptualized by information-seeking models in Section~\ref{sec:hypotheses:models}. Section~\ref{sec:hypotheses:activities} details how the cognitive and affective activities in these stages have been studied in the literature and accordingly defines our hypotheses. Finally, Section~\ref{sec:hypotheses:signals} discusses how physiological signals and the derived indexes can be used to characterize cognitive load, affective arousal, and affective valence.

\subsection{Information Seeking Models}
\label{sec:hypotheses:models}
Several information-seeking models have been proposed in the literature \cite{kuhlthau2005information, cole2011theory,marchionini1995information,belkin1980anomalous}. Similarly to \citet{moshfeghi2018search}, we characterize the informational process with a sequence of search stages that reflect a consensus among these models: Realization of Information Need (\ifn), Query Formulation (\qf), Query Submission (\qs), and Relevance Judgment (\rj).\footnote{Satisfaction Judgment is not considered in this paper; to reduce complexity and possible confounding variables, we only use a single result item during the RJ process (rather than reading a SERP that presents a ranking of items, or a session).}
Note that this theoretical framework presented here is our adaptation of a handful of former models that, we view, were incomplete. Hence, it requires an amalgamation of former theories and unities, as shown in Figure~\ref{fig:flow_chart}.

\begin{figure*}[tp]
    \centering
    \includegraphics[width=0.93\linewidth]{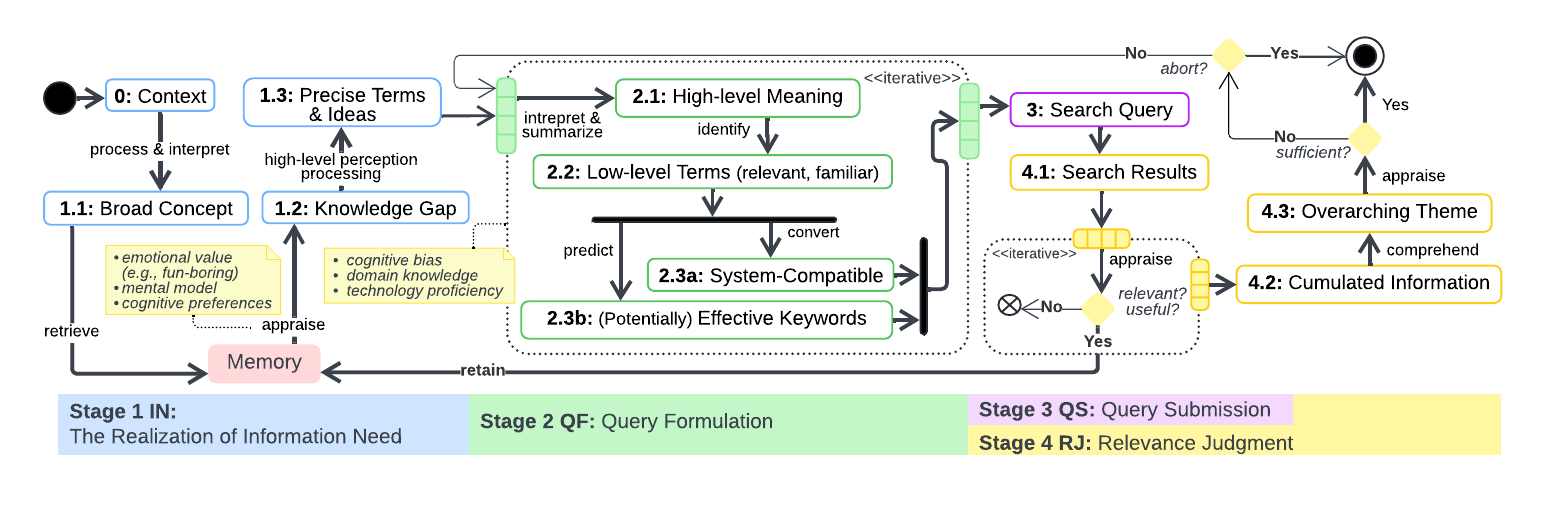}
    \setlength{\abovecaptionskip}{-5pt}
    \caption{The flow chart presents how the information is transformed through search stages, 1) the realization of Information Need (\ifn), 2) Query Formulation (\qf), 3) Query Submission (\qs) and 4) Relevance Judgment (\rj), in information seeking process, based on the combination and unification of the previous models.} 
    \Description{A colorful flowchart illustrating the stages of information seeking, from context and concept realization to query formulation, submission, and relevance judgment. Arrows connect these stages, and decision points are labeled throughout.}
    \label{fig:flow_chart}
\end{figure*}

\myparagraph{Realization of Information Need (\ifn)} See `Stage 1' with blue-colored borders presented in Figure~\ref{fig:flow_chart}. Users start with a `vague' idea of the problem and gradually gain clarity \cite{kuhlthau2005information, moshfeghi2013cognition}.
Once information from external sources, such as visual or auditory channels, has been processed and understood \cite{moshfeghi2018search, allegretti2015relevance}, the next step involves retrieving relevant information from long-term memory, e.g., past experiences, learned concepts, and memories \cite{Dominika2022information, moshfeghi2019neuropsychological}, to articulate any knowledge gaps or informational needs \cite{cole2011theory, sutcliffe1998towards, savolainen2015interplay, belkin1980anomalous} (Stage 1.2 in Figure~\ref{fig:flow_chart}). Awareness is updated based on memory output \cite{Dominika2022information, moshfeghi2019neuropsychological}.
This is followed by high-level conceptualization \cite{sutcliffe1998towards, moshfeghi2019neuropsychological, nahl2007social, belkin1980anomalous} to refine the broad concepts into more specific and detailed terms and ideas \cite{sutcliffe1998towards, nahl2007social, cole2011theory} (Stage 1.3).

The outcome is a comprehensive framework that connects the specific details of an information need to a more extensive network of knowledge. This network includes background and contextual information along with related concepts. \citet{cole2011theory} refers to this framework as the ``Information Need Frame'' or ``broad focus'' as described by \citet{moshfeghi2019neuropsychological}. However, \citet{cole2011theory} also claims that the information need developed so far only scratches the surface. The deeper level requires several iterations of collection and refinement.
Mental models and cognitive preferences \cite{sutcliffe1998towards, nahl2007social, cole2011theory} might also steer the process, as in personalized understanding (e.g., filtering) and representation (e.g., organizing and structuring) of knowledge, and (emotional) value judgment\footnote{These variables and activities are also important at the \rj stage, as discussed later.} \cite{savolainen2015interplay}.

\myparagraph{Query Formulation (\qf)} See `Stage 2' with green-colored borders presented in Figure~\ref{fig:flow_chart}. 
Once the goal is clear, the initiative shifts from reactive (receiving information) to proactive (resolving uncertainty) \cite{savolainen2015interplay}. 
The desired outcome of \qf stage is a plan of action, specifically a strategy for obtaining useful information from the system.

To device that strategy, the searcher progressively accumulates internal information and knowledge about the topic matter to enhance the understanding of their foreground information need \cite{cole2011theory} (the background information need relates to distraction, see \citet{jiang2022understanding}).
Firstly, users interpret and create high-level meanings from the available information \cite{kuhlthau2005information, savolainen2015interplay}, mainly from memory or prior experience (Stage 2.1). Next, they identify lower-level terms that are relevant and familiar \cite{nahl2007social} (Stage 2.2) and convert into a language that is compatible with the system \cite{sutcliffe1998towards} (Stage 2.3a). They also predict which keywords will effectively lead to the desired information \cite{cole2011theory, kuhlthau2005information} (Stage 2.3b). Through multiple rounds of interpretation, identification, and prediction, the initial information need can connect with more specific and detailed needs \cite{sutcliffe1998towards, cole2011theory}. 
Here, users might also be influenced by their learned patterns of reasoning \cite{nahl2007social}, cognitive bias, and technology proficiency \cite{alaofi2022query}, to plan their search effectively \cite{savolainen2015interplay}.

\myparagraph{Query Submission (\qs)} See `Stage 3' with purple-colored borders presented in Figure~\ref{fig:flow_chart}. 
When the search query is ready, the next step is to express the query to the system and execute it \cite{moshfeghi2018search}. Modern systems offer various input modalities, such as typing via keyboard, speaking into a microphone, or more advanced approaches, such as, brain-computer interfaces \cite{eugster2016natural, ye2023relevance}.

\myparagraph{Relevance Judgment (\rj)} See `Stage 4' with yellow-colored borders presented in Figure~\ref{fig:flow_chart}. When search results are received, apart from comprehending and interpreting information like at \ifn  \cite{sutcliffe1998towards, allegretti2015relevance, moshfeghi2013understanding, moshfeghi2018search, pinkosova2022revisiting, paisalnan2021towards}, memory judgment~\cite{allegretti2015relevance} and inferential reasoning~\cite{ye2022towards, paisalnan2021towards} also apply to appraise the retrieved results \cite{moshfeghi2013understanding}. The criteria include relevance, usefulness, and sufficiency \cite{sutcliffe1998towards, nahl2007social}. 
If the information is deemed relevant, it will be retained in long-term memory \cite{pinkosova2022revisiting, ye2022towards} (from Stage 4.1 to 4.2). 
Moreover, \citet{cole2011theory} envisages the user beginning to recognize a broader picture beyond just the facts or data, but also the societal aspects behind the information need, such as problem-goal, problem-solution frameworks, or task formulas. \citet{moshfeghi2013understanding} and \citet{paisalnan2021towards} identified the brain regions that correspond to overarching theme comprehension activated at \rj (Stage 4.3).
Lastly, the appraisal outcomes influence the decision whether to continue the search \cite{nahl2007social}, either by adopting the results or modifying the search query \cite{sutcliffe1998towards}. 

\subsection{Cognitive \& Affective Activities in Search}
\label{sec:hypotheses:activities}

\emph{Cognitive load} refers to the amount of cognitive resources in working memory exerted to complete a task. Working memory, an important cognitive system in informational processing, is responsible for processing sensory information, controlling and coordinating cognitive resources, as well as caching and processing recalled memory \cite{gwizdka2010distribution, randall2014putting}. Within its finite capacity, the more the working memory is used, the better task performance can be achieved \cite{kumar2016measurement, chikhi2022eeg}. 
In terms of affective activities, \citeauthor{schubert1999measuring}'s model~\cite{schubert1999measuring} characterizes emotions with two main dimensions: (i) \emph{affective arousal}, which refers to the intensity of a feeling
and (ii) \emph{affective valence}, which refers to the direction (positive or negative) of the feeling 
\cite{savolainen2015interplay}.

Given the search stages identified above and these physiological constructs, we formalize phenomena observed in the literature with a set of hypotheses -- which we aim to test and validate with a laboratory user study.
We denote the hypotheses of \emph{cognitive load} as \hcog, \emph{arousal} as \haro, and \emph{valence} as \hval.

\myparagraph{Realization of Information Need (\ifn)}
This stage is about integrating information from the external context and internal memory. \ifn stage requires demanding cognitive effort~\cite{savolainen2015interplay} allocated for three important components, Memory Retrieval, Information Flow Regulation, and Decision-Making \cite{moshfeghi2019neuropsychological}. 

In an experiment, participants are usually given a set of backstories which simulate a scenario and evoke the need to search for information \cite{kelly2009methods}. 
A \textit{feeling of uncertainty} is elicited because of a knowledge gap \cite{kuhlthau2005information, moshfeghi2013cognition, belkin1980anomalous}, and might lead to a combination of negative feelings, such as irritation, confusion, frustration, anxiety, and rage \cite{savolainen2015interplay}. Even so, users still look forward to finding new information to solve their problems \cite{kuhlthau2005information}. 
A neurological experiment of \citet{moshfeghi2018search} encapsulates \ifn as a goal-setting process. Apart from the cognitive tasks for language processing, it also involves other tasks for which working memory is responsible, such as sustaining attention, planning, imagining, switching, maintaining instruction, and balancing and managing cognitive resources \cite{Dominika2022information, paisalnan2021towards, moshfeghi2019neuropsychological} (these are also involved in relevance judgment \cite{paisalnan2021towards, allegretti2015relevance}).
Subsequently, goal-directed feelings appear, which brings a sense of direction and temporary relief from negative feelings \cite{kuhlthau2005information, nahl2007social}. 
These anticipatory feelings and previous negative feelings might balance out \cite{savolainen2015interplay}. This explains the self-assessment results collected by \citet{moshfeghi2013cognition}, that participants experienced uncertainty, but low anxiety and neutral emotions were predominant. 
It also shows that these affective activities only hover at the subconscious level in practice, compared to the theory \cite{savolainen2015interplay}.

\myparagraph{Query Formulation (\qf)}
Now that the goal has become clear and a plan has been set, the initial feeling of uncertainty gradually decreases while confidence and clarity increase \cite{kuhlthau2005information, moshfeghi2013cognition}. It is progressing from planning to action, and the users are ready to begin to search \cite{kuhlthau2005information}. Participants in \citet{moshfeghi2018search} and \citet{shovon2015search}'s experiments are also found to be prepared and ready to express at this stage. The cognitive activities here mainly involve term interpretation, identification, and prediction. 
We therefore expect the following relationships:
$$
\resizebox{\linewidth}{!}{
$\begin{array}{rcl}
\multicolumn{3}{l}{\emph{\ifnmath \mbox{ versus } \rjmath \mbox{ :}}} \\
\hcogmath(1): & \mbox{cognitive load}(\ifnmath) > \mbox{cognitive load}(\qfmath) & 
\text{\cite{moshfeghi2019neuropsychological, Dominika2022information, paisalnan2021towards, moshfeghi2018search}}\\
\haromath(1): & \mbox{arousal}(\ifnmath) > \mbox{arousal}(\qfmath)  & \text{\cite{moshfeghi2019neuropsychological, Dominika2022information, paisalnan2021towards, kuhlthau2005information}}\\
\hvalmath(1): & \mbox{valence}(\ifnmath) < \mbox{valence}(\qfmath) & \text{\cite{moshfeghi2013cognition, nahl2007social, paisalnan2021towards, kuhlthau2005information}}
\end{array}$
}
$$

\myparagraph{Query Submission (\qs)}
Both \citet{gwizdka2010distribution} and \citet{shovon2015search} observed that formulating and submitting queries requires more cognitive effort than passively receiving information (in relevance judgment). They reasoned that this is due to the simultaneous cognitive processes involved in recalling and producing terms being more demanding. 
The findings of \citet{moshfeghi2018search} differ from these two prior works, revealing that the brain activities at \qf are primarily associated with semantic interpretation, keyword identifications and formulation, and prediction. At \qs, they are centered around attention and motor processing for expressing (verbalizing) the query, and affective activities related to reward processing. We therefore expect:
$$
\resizebox{\linewidth}{!}{
$\begin{array}{rcl}
\multicolumn{3}{l}{\emph{\qfmath \mbox{ versus } \qsmath \mbox{ :}}} \\
\hcogmath(2): & \mbox{cognitive load}(\qfmath) < \mbox{cognitive load}(\qsmath) & \text{\cite{moshfeghi2018search}}\\
\\
\multicolumn{3}{l}{\emph{\qsmath \mbox{ versus } \rjmath \mbox{ :}}} \\
\hcogmath(3): & \mbox{cognitive load}(\qsmath) > \mbox{cognitive load}(\rjmath) & \text{\cite{paisalnan2021towards, moshfeghi2013understanding, allegretti2015relevance, moshfeghi2018search}}
\end{array}$}
$$

For the affective activities, \citet{savolainen2015interplay} supposes that feelings are combined with positives related to brief elation and anticipation, and negatives such as confusion and sometimes anxiety, at \qs. 
However, \citet{lopatovska2014toward} found no significant variation of facial expressions during \qs, yet collected insufficient self-rating emotion data. Explicitly, \citet{moshfeghi2018search} found the brain regions responsible for affective appraisal activate at \qs, confirming the occurrences of affective activities. This also implies that emotions are triggered by the expectation of the query's possible success and the inherent reward of finding the right information. Taken together, these results align with \citeauthor{savolainen2015interplay}'s discussion \cite{savolainen2015interplay}, a balance between anticipatory emotions and overall emotional tone, so that most affective activities stay at a subconscious level.
Furthermore, the difference between actively expressing at \qs and passively receiving at \qf might distinguish the arousal level.
We expect:
$$
\resizebox{0.7\linewidth}{!}{
$ 
\begin{array}{rcl}
\multicolumn{3}{l}{\emph{\qfmath \mbox{ versus } \qsmath \mbox{ :}}} \\
\haromath(2): & \mbox{arousal}(\qfmath) < \mbox{arousal}(\qsmath) & \text{\cite{kuhlthau2005information}}\\
\hvalmath(2): & \mbox{valence}(\qfmath) < \mbox{valence}(\qsmath) & \text{\cite{kuhlthau2005information}}
\end{array}$
}
$$

\myparagraph{Relevance Judgment (\rj)}
As depicted by \citet{kuhlthau2005information}, when the search process nears completion,  feelings generally shift to predominantly positive. The level of uncertainty decreases, and users feel more confident as they become better at finding relevant information. Interest also increases. In particular, users often experience satisfaction and a sense of direction when they come across useful information, as they can navigate through the information more effectively. Conversely, if the information is not useful, boredom can set in. This theory was later supported in experiments, such as self-reported perception from \citet{moshfeghi2013cognition}, anticipatory electrodermal responses from \citet{mooney2006investigating}, and increasing sadness when search results fail to meet expectations from \citet{lopatovska2014toward, arapakis2008affective}. 
In particular, self-assessment collects less neutral emotion \cite{moshfeghi2013cognition}, and the most frequent facial expression is surprise \cite{lopatovska2014toward, arapakis2008affective, mcduff2021affective, moshfeghi2013cognition}. The results captured by these approaches mean that feelings reach a conscious level, indicating the intensity of feelings is stronger at \rj.
It is worth noting that high cognitive load is usually associated with high arousal \cite{hogervorst2014combining}, but not solely. 
Both cognitive and affective perspectives suggest \rj has the highest level of arousal. 
Therefore: 
$$ 
\resizebox{0.7\linewidth}{!}{$
\begin{array}{rcl}
\multicolumn{3}{l}{\emph{\qsmath \mbox{ versus } \rjmath \mbox{ :}}} \\
\haromath(3): & \mbox{arousal}(\qsmath) < \mbox{arousal}(\rjmath) & \text{\cite{moshfeghi2018search, lopatovska2014toward}}\\
\hvalmath(3): & \mbox{valence}(\qsmath) < \mbox{valence}(\rjmath) & \text{\cite{kuhlthau2005information}}
\end{array}$}
$$

Moreover, although \ifn and \rj both involve passively receiving information, the latter requires more demanding cognitive processes \cite{paisalnan2021towards}. The efforts at \rj mainly are exerted to encode and maintain the task (e.g., relevance criteria), store and update information, and accumulate evidence during appraisal \cite{paisalnan2021towards, allegretti2015relevance, moshfeghi2013understanding}. 
Meanwhile, negative feelings can still dominate at \rj, reflecting unsatisfying results, or greater mental effort or concentration when dealing with challenges like information overload, conflicts, or complex information \cite{kuhlthau2005information, savolainen2015interplay}; the results of \citet{mcduff2021affective} and \citet{gwizdka2010distribution} provide observational support. 
Accordingly, we expect: 
$$
\resizebox{\linewidth}{!}{
$\begin{array}{rcl}
\multicolumn{3}{l}{\emph{\rjmath \mbox{ versus } \ifnmath \mbox{ :}}} \\
\hcogmath(4): & \mbox{cognitive load}(\rjmath) > \mbox{cognitive load}(\ifnmath) & \text{\cite{paisalnan2021towards}} \\
\haromath(4): & \mbox{arousal}(\rjmath) > \mbox{arousal}(\ifnmath) & \text{\cite{moshfeghi2013cognition, mooney2006investigating}} \\
\hvalmath(4): & \mbox{valence}(\rjmath) > \mbox{valence}(\ifnmath) & \text{\cite{arapakis2008affective, lopatovska2014toward, moshfeghi2013cognition, allegretti2015relevance, mcduff2021affective}}
\end{array}
$}
$$

\subsection{Physiological Indexes}
\label{sec:hypotheses:signals}
Physiological indexes are measurable biological functions that provide insights into an individual's activities, such as their physical and emotional state, cognitive performance, and overall health. 

\myparagraph{Cognitive Load} 
The different intensities of signals generated by the human brain can indicate various cognitive activities.
The frontal cortex plays a crucial role in attention, memory, and judgment. 
A common agreement that the frontal theta power (4--8 Hz) is a strong indicator for the change of cognitive load \cite{chikhi2022eeg, puma2018using}, regardless of visual or auditory modalities \cite{kaminski2016information} or cognitive or motor type of tasks \cite{so2017evaluation}. Increased cognitive load is associated with enhanced frontal theta.
Another brain wave, alpha power (8--12 Hz) is also frequently mentioned in relation to measuring cognitive load. Alpha power predominates when relaxing or inhibiting task-irrelevant activities \cite{raufi2022evaluation, puma2018using, chikhi2022eeg}. Although there are some inconsistent results, \citet{chikhi2022eeg} synthesizes the existing findings and reveals a prevalent negative correlation of cognitive load on alpha power in the parietal cortex -- responsible for sensory processing. 
Combining these two, Theta-Alpha Ratio (TAR) has been validated by \citet{raufi2022evaluation} as an index level of cognitive load.

Pupil dilation is also extensively used to measure cognitive load \cite{Gwizdka2017temporal, van2018pupil}, with increasing cognitive load being associated with increasing pupil dilation \cite{van2018pupil, puma2018using}. Compared to the highly-sensitive nature of EEG with multiple channels, pupil data can provide a cleaner and simpler indication of cognitive load \cite{puma2018using}. 

Both TAR and pupil dilation are typically positively correlated with cognitive load, but they contribute from different aspects. Pupil dilation is usually associated with the attentional aspect of cognitive load or general affective arousal \cite{Gwizdka2017temporal, puma2018using, van2018pupil, gwizdka2018inferring}, whereas TAR is more specifically tied to the intensity of neural activity when engaged in cognitive or memory processing tasks \cite{sauseng2002interplay, raufi2022evaluation}.

\myparagraph{Arousal}
Apart from theta and alpha, beta power (12-30 Hz) is also influenced by cognitive load. But it is caused by an associative relationship from emotional responses or other underlying mechanisms \cite{chikhi2022eeg}; enhanced cognitive load might associate with enhanced affective arousal \cite{hogervorst2014combining}. Beta power is associated with an alert or excited state of mind \cite{ramirez2012detecting}, while alpha power is associated with a relaxed state. They are often used as a robust index of \textit{arousal}, computed as the Beta-Alpha Ratio (BAR) \cite{matlovivc2016emotion, ramirez2012detecting}. When experiencing high arousal, the level of beta should be high while alpha should be low, resulting in a high BAR \cite{matlovivc2016emotion}.

In addition, Electrodermal Activity (EDA) and Photoplethysmogram (PPG) are also robust indicators of arousal. Specifically, high arousal elicits presentation in higher Skin Conductance Level (SCL) \cite{eda2021baba, greco2016arousal} and Heart Rate Variability (HRV) \cite{hogervorst2014combining, bias2023boon, pham2021heart}. As mentioned above, \citet{mooney2006investigating} has found increased EDA at \rj, indicating anticipatory feelings.

\myparagraph{Valence} 
It is widely accepted in psychological studies that alpha power between the left and right frontal areas is associated with emotion \cite{harmon2003clarifying, lee2020frontal, harmon2018role}. In particular, enhanced left alpha is associated with negative emotion or withdrawal response, and vice versa. Regarding information activity, this withdrawal/approach response is represented as being open or conservative towards new information \cite{kuhlthau2005information, savolainen2015interplay}. 
Therefore, the level of asymmetry of alpha power in the frontal area, Frontal Alpha Asymmetry (FAA), is usually used to measure valence \cite{matlovivc2016emotion, ramirez2012detecting, harmon2018role}. A negative FAA indicates relatively higher left alpha, thus negative emotion.

\section{User Study}
\label{sec:experiment}

\begin{figure}[tp]
    \centering
    \includegraphics[width=0.9\linewidth]{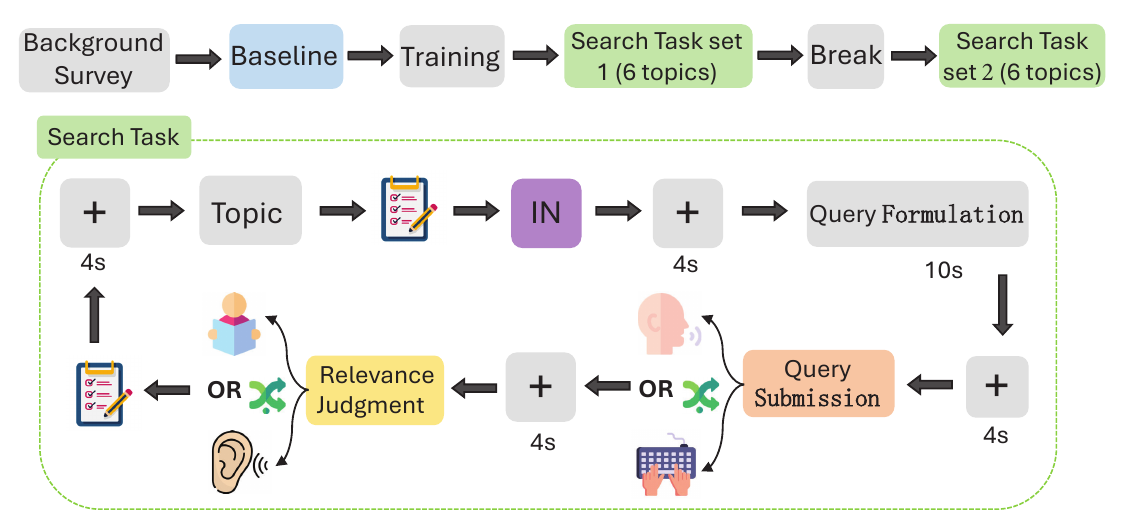}
    \caption{Experimental Procedure. \ifn: the realization of Information Need.}
    \Description{A top flowchart illustrates the steps of the experiment, including Background Survey, Baseline, Training, Search Tasks Set 1 (contains six topics), and Break, then Search Tasks Set 2 (contains another six topics). The bottom flowchart illustrates the steps of the search task. }
    \label{fig:exp1fig}
\end{figure}

\subsection{Procedure}

The experimental protocol is shown in Figure~\ref{fig:exp1fig}\footnote{Materials and code are available at \url{https://github.com/kkkkk2017/IR-Physiological-Signals}. The data can be requested by contacting the authors.}. 
After calibration, the participants answer a background survey; information about handedness, sleep quality and caffeine intake are collected. Next, the participants complete a 15-second eyes-open (\eyeopen) and a 15-second eyes-closed (\eyeclose) section to collect the baseline data, followed by a training section containing the instruction and two practice tasks. Then they proceed to perform the search tasks (12 in total). 

For the search task, participants start by looking at a fixed cross in the middle of a blank screen for 4 seconds. Next, a topic title is shown. Then, participants rate their interest, familiarity, and expected difficulty regarding the topic using a 5-level Likert item. Next, a backstory that evokes the information need (\ifn) is presented. Participants are then given 10 seconds to form a search query in their mind (\qf), followed by submitting the query (\qs) either written in text or via voice. Once the query is submitted, participants receive one relevant information snippet -- either displayed as text on the screen, or played as an audio clip. Finally, they need to answer a binary factual judgment question (attention check) and rate their perceived relevance and difficulty in understanding the search result. In order to account for the delays on physiological responses a 4-second fixed cross gap is provided between search stages, i.e., \ifn, \qf, \qs, and \rj.  

The sequences of topics and the interaction modalities (voice or text) are randomized. 
A mandatory 5-minute break is taken after 6 tasks. After completing the search tasks,
participants verbally describe their experiences towards the experiment for quality purposes.
Furthermore, to capture the activities precisely, we record all the timestamps of page transactions, bottom press (to start/stop voice input), and first and last keystroke input.

\subsection{Materials}
\myparagraph{Information Needs} We use the backstories in the InformationNeeds dataset \cite{bailey2014information} created by \citet{informationneed2014}. The dataset contains backstories that represent different information needs for 180 TREC topics. The information needs were categorized into three levels of cognitive complexity: Remember, Understand, and Analyze. We choose 12 topics
from the middle level (Understand) to have enough room for unfamiliarity, but also to avoid risks of triggering emotions or cognitive bias. The Understand category involves searching and gathering relevant messages to construct meaning for the given topic. We randomly sample topics and remove those related to crises, wars, conspiracy, or politically sensitive topics. The original backstories have an average of 41 words ($SD=6$). To ensure all selected backstories have a similar word count, we manually edited them, resulting in an average of 40 words ($SD=1$). 

\myparagraph{Search Results} For each information need, participants receive one information snippet generated by combining relevant documents as follows. 
Although the backstories~\cite{informationneed2014} were developed based on TREC topics, the qrels from the corresponding TREC test collection does not directly align with the Information Need. Therefore, 
given the TREC topic associated with the backstory, we manually select up to three documents judged as relevant in the qrels. Then, we use GPT-3.5\footnote{\url{https://chat.openai.com/}} to generate a 150-word summary based on the provided documents\footnote{The questions are generated using the prompt below: ``Based on these articles, can you write me a 150-word summary to tell me [backstory] [relevant documents]''} -- as well as a binary factual judgment question that we used as attention check.
To minimize the influences of word lengths or complexity, we further manually examine the generated summaries using the Flesch Reading Ease (FRE) score \cite{flesch1948new}. Overall, the summaries have an average word count of 148 ($SD=3$) and an average FRE score of 11.9 ($SD=0.9$).

\subsection{Equipment and Setup}

\begin{figure}[htbp!]
    \centering
    \begin{subfigure}[t]{.6\linewidth}
        \centering
        \includegraphics[width=\linewidth, trim={0 5cm 0 0}, clip]{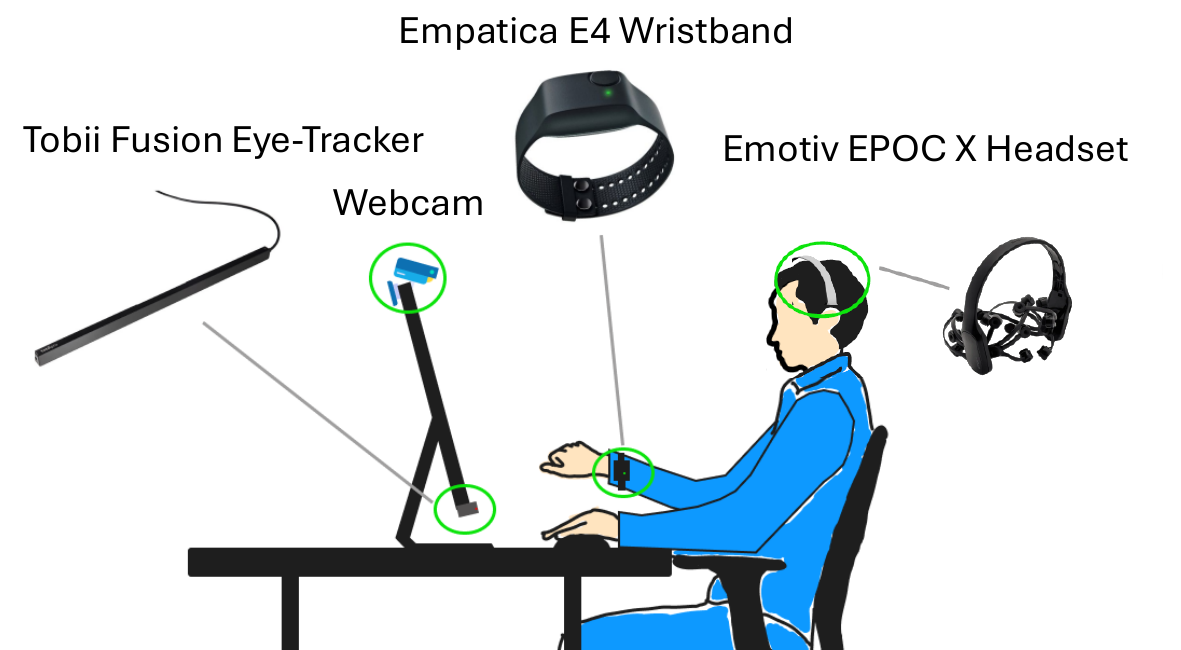}
    \end{subfigure}
    \begin{subfigure}[t]{.3\linewidth}
        \centering
        \includegraphics[width=\linewidth, trim={8mm 2mm 5mm 0}, clip]{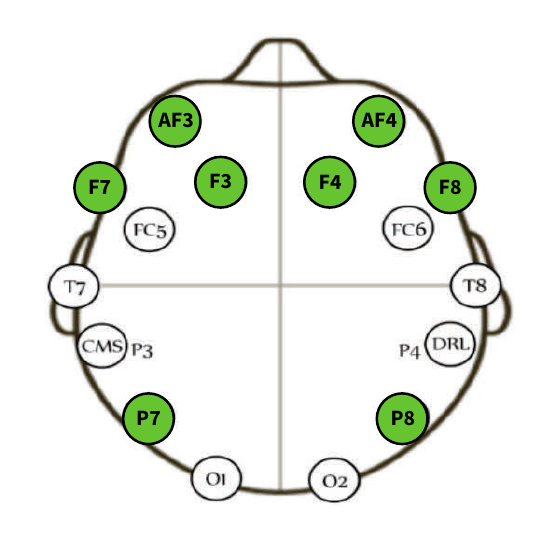}
    \end{subfigure}
        \caption{Experiment setup (left), and the EEG electrode locations (right). The filled circles indicate the electrodes used to compute the indexes.}
        \Description{The left figure illustrates a person sitting in front of a computer and wearing various devices for data collection. The right figure depicts EEG electrode placements on a head.}
        \label{fig:setup}
\end{figure}

Four sensors are used in this study: a webcam camera for video recording, a Tobii Fusion eye-tracker\footnote{\url{https://www.tobii.com/products/eye-trackers/screen-based/tobii-pro-fusion}} for pupillary responses (60Hz), an E4 wristband\footnote{\url{https://www.empatica.com/en-int/research/e4/}} for EDA (4Hz) and PPG (64Hz), and a 14-channel Emotiv EPOC headset\footnote{\url{https://www.emotiv.com/epoc-x/}} for EEG data (128Hz). 
The experiment is conducted in an illuminated room. The participant sits in front of a desktop PC, which is mounted with an eye-tracker and a web camera. 
All participants use the computer mouse with their right hand, and wear the wristband on the left hand. We sanitize the electrodes and the participant's skin on the inner and outer wrist with alcohol wipes \cite{eda2021baba}. Then, the instructor helps the participants to wear the headset, and adjusts the positions of the electrodes. The experiment material is deployed using the Qualtrics\footnote{\url{https://www.qualtrics.com/about/}} platform.

\subsection{Participants}

The study received human research ethics approval from RMIT University, and participants provided written informed consent prior to the experiment. To ensure a minimum of additional effort involved for language, we recruit participants with at least a professional working proficiency level in English. A total of 29 participants are recruited. The data collected for 3 of these participants are discarded due to environmental disturbances. Due to software errors, the eye-tracking data from 3 participants could not be obtained. For results concerning EEG, EDA, and PPG, we use valid data from 26 participants (15M, 11F). There are 77\% of the participants with full professional proficiency or are native English speakers. For results related to eye data, we use valid 23 participants (13M,10F).

\section{Data Clean-Up \& Analysis}

First, the data obtained from all sensors are synchronized by timestamp. Each recording is then denoised, explained in further detail below, and divided into 13 trials corresponding to 1 baseline (EYEOPEN/CLOSE) and 12 search tasks. As per our experimental methodology, each trial starts with a 4-second fixation, and contains 4 Events of Interest (EOI), i.e., \ifn, \qf, \qs, \rj. To deal with time inconsistency, we only analyze the first 10 seconds of each EOI, selected by the lower quartile \cite{so2017evaluation}.

\begin{table}[tb]
\caption{Data cleanup summary. Note that each baseline (in parentheses) corresponds to data from one participant.}
\label{tab:data_clean_summ}
\resizebox{\linewidth}{!}{
\begin{tabular}{lllcccc}
    \toprule
    \multicolumn{3}{l}{\multirow{2}{*}{Data cleanup step}}       & \multicolumn{4}{c}{ \underline{Number of Trials (+ Baseline)}} \\ 
                                 & &              & \multicolumn{2}{c}{EEG \& EDA \& PPG}  & \multicolumn{2}{c}{PUPIL}      \\ \midrule
    \multicolumn{3}{l}{Original data}                              & \multicolumn{2}{c}{312 (+26)}   & \multicolumn{2}{c}{276 (+23)}  \\
    \multicolumn{3}{l}{Bad data cleanup}                          & \multicolumn{2}{c}{300 (+25)}   & \multicolumn{2}{c}{182 (+23)}  \\
    \multicolumn{3}{l}{Removal by self-ratings}                    & \multicolumn{2}{c}{177 (+25)} & \multicolumn{2}{c}{159 (+23)}  \\
    \multicolumn{3}{l}{Removal of 1 person with only 1 trial} & \multicolumn{2}{c}{176 (+24)}   & \multicolumn{2}{c}{158 (+22)}  \\ 
    \bottomrule
\end{tabular}
}
\end{table}

\subsection{Data Processing}
\textbf{EEG} data is processed using the MNE Python library.\footnote{\url{https://mne.tools/stable/i}} The break section is excluded. 
Following similar procedures to \citet{martinez2023impact} and \citet{Gwizdka2017temporal}, each EEG recording is first denoised with a Butterworth filter (1--50Hz, $5^{th}$), removed the signal mean, 
and re-referenced with the common average. 
Next, the data is further cleaned and interpolated with the Autoreject \cite{jas2017autoreject} package. Lastly, to remove the artifacts (e.g., blinking), we use the Independent Component Analysis (ICA) combined with ICLabel \cite{Li2022}.
One recording is removed because of bad quality of EEG data.
The power spectral density of each EEG channel is then calculated using Welch's method and hamming window and normalized \cite{kosonogov2023eeg, so2017evaluation, lee2020frontal}.
The indexes are then computed from the set of EEG electrodes as follows.
\textbf{Theta-Alpha Ratio (TAR)} is computed by
\(avg(\theta(AF3, AF4, F3, F4, F7, F8))/avg(\alpha(P7, P8))\) \cite{raufi2022evaluation}.
\textbf{Beta-Alpha Ratio (BAR)} and \textbf{Frontal Alpha Asymmetry (FAA)} are computed \( BAR = \beta(AF3 + AF4 + F3 + F4)/\alpha(AF3 + AF4 + F3 + F4)\),
\( FAA = \log(\alpha(F4)/\beta(F4)) - \log(\alpha(F3)/\beta(F3))\)  \cite{matlovivc2016emotion, ramirez2012detecting, harmon2018role}.

\textbf{Pupil} data are cleaned following the procedure described by \citet{kret2019preprocessing} and \citet{Gwizdka2017temporal}. The left and right pupils are first processed separately. Samples with dilation speed above the median absolute deviation or the gap between two data points above (75 ms) are removed. This is done twice for each side to remove the edge values. Then, the cleaned data of both sides are combined by taking the arithmetic mean, and linear interpolation is applied to fill in the blink gaps. Finally, a zero-phase Butterworth filter (4Hz, $3^{rd}$) is applied to remove outliers \cite{Martin2021}.
As our experiment includes sub-tasks that do not require on-screen visuals (i.e., \qf via voice and \rj via audio), some sub-tasks are significantly lacking in pupil data. The EOIs with > 20\% missing data are excluded for analysis \cite{Gwizdka2017temporal}, and the trials that do not include all 4 EOIs are subsequently excluded.  
\textbf{Relative Pupil Dilation (RPD)} calculates the relative changes of current pupil diameter compared to a baseline value \cite{Gwizdka2017temporal}: 
\(RPD_{t}^{i} = (P_{t}-P_{baseline}^{i})/P_{baseline}^{i}\),
where $t$ is time, $i$ is participant, and baseline is the average pupil diameter across all tasks.

\textbf{EDA and PPG} signals obtained from the wristband are processed using the NeuroKit2~\cite{makowski2021neurokit} Python library, following a similar procedure as by \citet{Lascio2018student, Bota2019, braithwaite2013guide}. For EDA, a low-pass (0.5Hz) Butterworth filter followed by a rolling median with a 3-second window \cite{eda2021baba} and min-max normalization are applied. The convex optimization \texttt{cvxEDA} method \cite{greco2016cvxeda} is then applied to decompose the tonic value, i.e., the \textbf{Skin Conductance Level (SCL)}.
The raw PPG data is cleaned with the default approach in NeuroKit2. Then, the time between consecutive heartbeats is computed, representing \textbf{Heart Rate Variability (HRV)} in milliseconds.

\subsection{Assumptions \& Trial Selection}

When forming the hypotheses, it is worth noting that the following assumptions
are made when considering possible factors that might interfere with physiological responses, such as information complexity \cite{martinez2023impact}, relevance \cite{ye2023relevance, allegretti2015relevance, eugster2016natural, oliveira2009discriminating, barral2015exploring}, and interest \cite{Wise2009, White2017881}. 

In this experiment, the participants report average scores of 3.5 ($SD=1.1$) interest, 2.5 ($SD=1.1$) difficulty, 2.6 ($SD=1.3$) familiarity towards the topics, and  4.0 ($SD=1.1$) relevance, 2.0 ($SD=1.1$) difficulty to the search results. To meet the assumption before conducting any analysis, we first select from the trials based on self-ratings, using the following thresholds:

\begin{itemize}
    \item Users are fairly interested in the topics (1 \textless{} topic\_interest \textless{} 5). 
    \item The search results are relevant to the submitted queries (info\_\\relevance $\geq 3$).
    \item Search results are not difficult to understand (info\_difficulty $\leq 3$). 
    \item Participants are engaged in the tasks.
\end{itemize}

The summary of data cleanup is presented in Table~\ref{tab:data_clean_summ}.
It is noteworthy that \qf and \qs are usually consecutive phases. Although our experimental design attempts to separate them, it cannot guarantee the complete removal of automatic progression. It might also involve \qf-related cognitive activities at \qs, such as recalling and re-evaluating the terms.

\subsection{Statistical Analysis}
The proposed hypotheses are tested in a within-subject setting.
As the data is not normally distributed, we conduct the non-parametric Wilcoxon signed-rank tests for each physiological index between pairs of EOIs: \ifn and \qf, \qf and \qs, \qs and \rj, \rj and \ifn. The Bonferroni correction for multiple comparisons is applied to adjust p-values before they are compared to the $\alpha$ significance thresholds.

\section{Results}

\begin{table}[tb]
\caption{Summary of hypothesis validation. Pairs with 
significant differences that confirm the hypothesis 
({\color[HTML]{009901}\cmark})
, a significant but opposite relationship
({\color[HTML]{FE0000}\xmark})
or no significant difference 
({\color[HTML]{FE0000} --})
. $p$ < .001***, $p$ < .01**, $p$< .05*.
}
\resizebox{\linewidth}{!}{
\begin{tabular}{ccc|cccc|cc}

\toprule
   \hcog    & \textbf{TAR}                            & \textbf{RPD}        &    \haro     & \textbf{BAR}                            & \textbf{SCL}& \textbf{HRV}                            &     \hval    & \textbf{FAA} \\ \midrule
   
{  IN \textgreater{} QF} &
{\color[HTML]{009901}\usym{1F5F8}}***& {\color[HTML]{FE0000} \usym{2718}}***&
{  IN \textgreater{} QF} & 
{\color[HTML]{009901}         \usym{1F5F8}}* & {\color[HTML]{FE0000} --}                & {\color[HTML]{FE0000} --}                &
{  IN \textless{} QF}& 
{\color[HTML]{FE0000} --} \\

{  QF \textless{} QS}        & 
{\color[HTML]{009901}         \usym{1F5F8}}***& {\color[HTML]{009901}         \usym{1F5F8}}***& 
{  QF \textless{} QS}        & 
{\color[HTML]{FE0000} --}                & {\color[HTML]{FE0000} --}     & {\color[HTML]{FE0000} \usym{2718}}***&
{  QF \textless{} QS}        &
{\color[HTML]{FE0000} --} \\

{  QS \textgreater{} RJ}     &
{\color[HTML]{009901}         \usym{1F5F8}}***& {\color[HTML]{009901}         \usym{1F5F8}}***& 
{  QS \textless{} RJ}        & 
{\color[HTML]{FE0000} --}    & {\color[HTML]{009901}         \usym{1F5F8}}* & {\color[HTML]{009901}         \usym{1F5F8}}***& 
{  QS \textless{} RJ}        & {\color[HTML]{FE0000} --} \\

{  RJ \textgreater{} IN }     &  
{\color[HTML]{FE0000} --}                & {\color[HTML]{009901} \usym{1F5F8}}***& 

{  RJ \textgreater{} IN}     &
{\color[HTML]{FE0000} --}                & {\color[HTML]{FE0000} --}                & {\color[HTML]{009901}  \usym{1F5F8}}** & 

{  RJ \textgreater{} IN }     & 
{\color[HTML]{FE0000} --} \\ \bottomrule
\end{tabular}}

\label{tab:hyp_res_summ}
\end{table}

\begin{figure}[tb]
    \centering
    \begin{subfigure}[b]{.45\linewidth}
        \includegraphics[width=\linewidth]{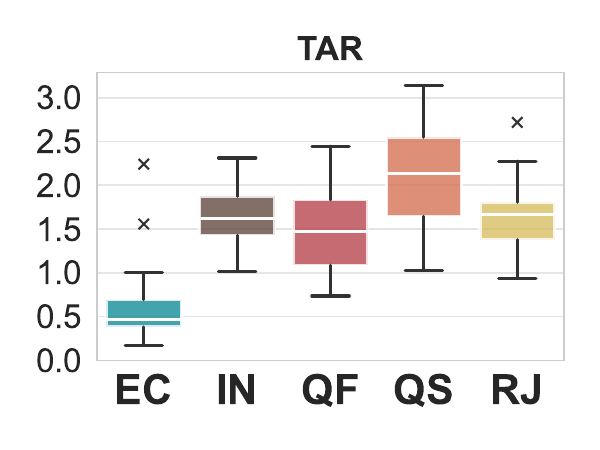}
    \end{subfigure}
    \begin{subfigure}[b]{.45\linewidth}
        \includegraphics[width=\linewidth]{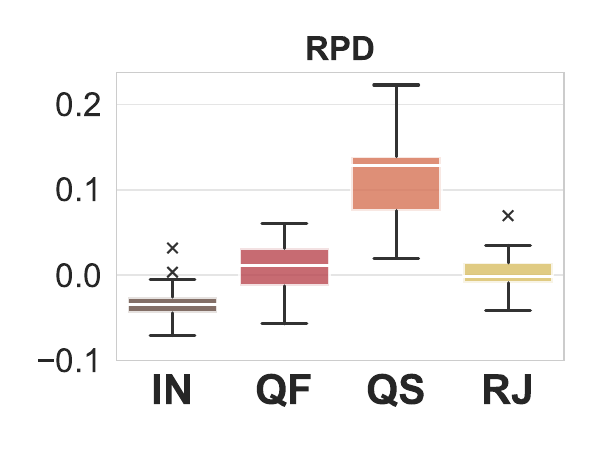}
    \end{subfigure}
    \caption{Distribution of indexes for measuring \emph{cognitive load} across all participants. The values of one participant are aggregated into one data point. }
    \Description{The left box plot represents ``TAR" related to 5 categories (EC, IN, QF, QS, RJ). The right box plot represents ``RPD" related to 4 categories (IN, QF, QS, RJ). }
    \label{fig:cogn_results}
\end{figure}

\begin{figure*}[tb]
    \centering
    \begin{subfigure}[b]{.2\linewidth}
    \includegraphics[width=\linewidth]{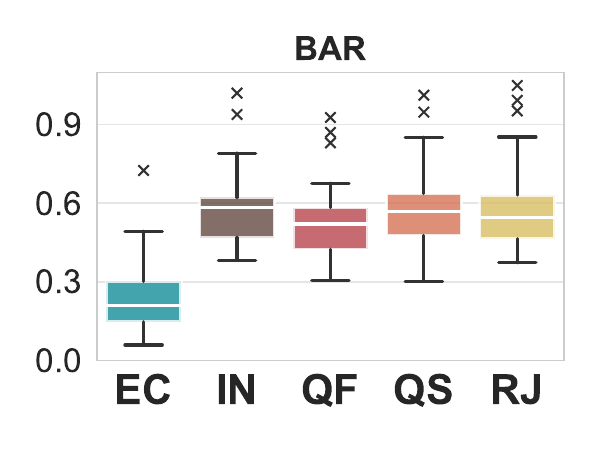}
        \caption{} 
        \label{subfig:BAR}
    \end{subfigure}
    \begin{subfigure}[b]{.2\linewidth}
        \includegraphics[width=\linewidth]{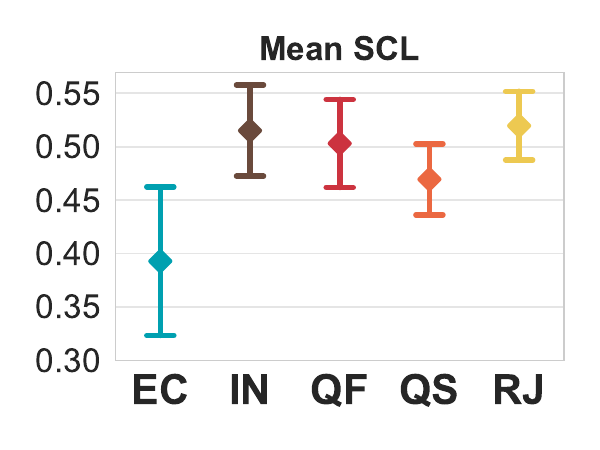}
        \caption{}
        \label{subfig:SCL}
    \end{subfigure}
    \begin{subfigure}[b]{.2\linewidth}
        \includegraphics[width=\linewidth]{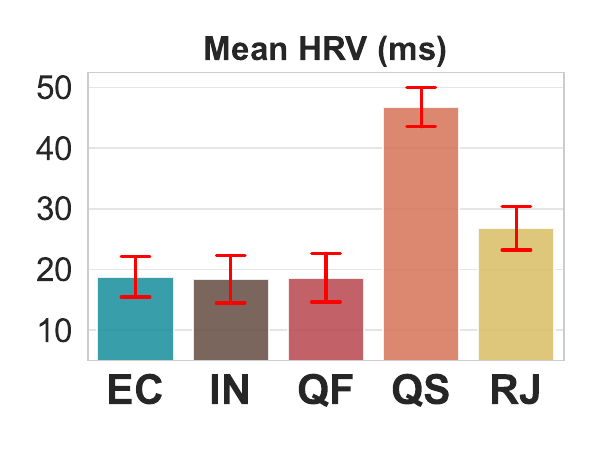}
        \caption{}
        \label{subfig:HRV}
    \end{subfigure}
    \begin{subfigure}[b]{.203\linewidth}
        \includegraphics[width=\linewidth]{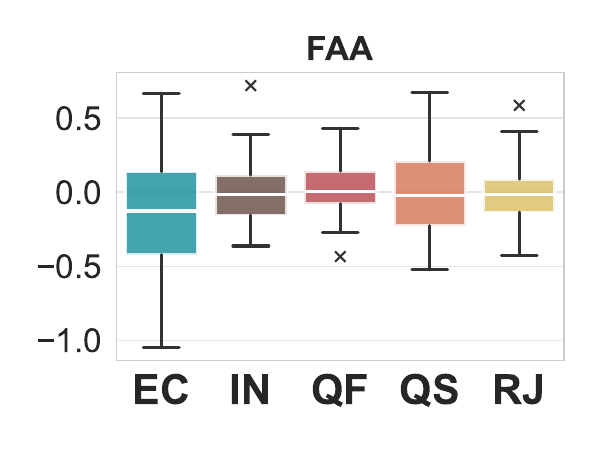}
        \caption{}
        \label{subfig:FAA}
    \end{subfigure}
    \begin{subfigure}[b]{.13\linewidth}
        \includegraphics[width=\linewidth]{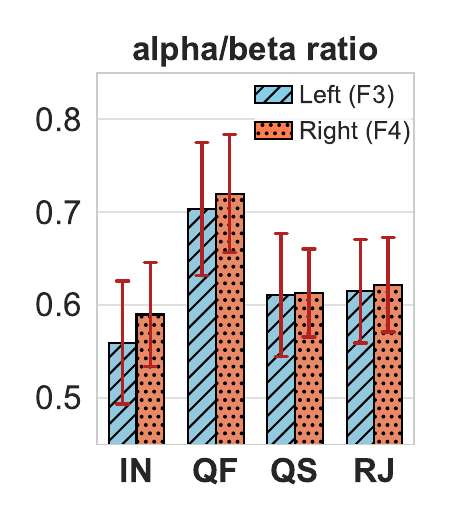}
        \caption{}
        \label{subfig:FAA_components}
    \end{subfigure}
    \caption{Distribution of indexes for measuring \emph{affective arousal} (\protect\subref{subfig:BAR}, \protect\subref{subfig:SCL}, \protect\subref{subfig:HRV}) and \emph{valence} (\protect\subref{subfig:FAA}, \protect\subref{subfig:FAA_components}). Error bars indicate standard error.}
    \label{fig:arousal_valence_result}
    \setlength{\belowcaptionskip}{-1pt}
    \Description{The figure contains 5 different plots representing each index related to 5 categories (EC, IN, QF, QS, RJ). Figure A, B and D are box plots, while Figure C and E are bar plots.}
\end{figure*}

\subsection{Baseline} 
\textit{EYECLOSE} (\eyeclose) represents a relaxed state of participants, potentially indicating a minimum level of cognitive effort, arousal, and valence. 
However, \eyeclose is excluded when comparing RPD as pupil data is unavailable.
TAR and BAR have significant differences at \eyeclose compared to all EOIs, but FAA or SCL does not. 
Nevertheless, SCL is lower than all EOIs (refer to Figure~\ref{subfig:SCL}).
HRV has significant differences at \eyeclose compared to \qs or \rj.

\subsection{Cognitive Load (\hcog)}
\label{subsec:result_cog}

\paragraph{\hcog(1): \ifn versus \qf}
Both TAR and RPD show significant differences between \ifn and \qf ($W_{TAR}=28, p<.001$, $W_{RPD}=20, p<.001$). But interestingly, they present opposite trends, as shown in Figure~\ref{fig:cogn_results}.  
TAR is higher at \ifn than \qf, whereas RPD is lower. These opposite results can be presumably explained by the different cognitive demands required at these EOIs. As discussed in Section~\ref{sec:hypotheses}, the primary cognitive activities at \ifn involve information processing and memory retrieval; thus, higher TAR. 
In contrast, those at \qf primarily entail problem-solving to generate an effective search query, attentional resources are dominant; thus, higher RPD.
Overall, \hcog(1) is partially supported.

\paragraph{\hcog(2): \qf versus \qs, and \hcog(3): \qs versus \rj}
Both TAR and RPD are significantly lower at \qf than \qs ($W_{TAR}=1, p<.001$, $W_{RPD}=0, p<.001$), which supports \hcog(2). They are also lower at \rj than \qs ($W_{TAR}=3, p<.001$, $W_{RPD}=1, p<.001$), which supports \hcog(3). 
These results suggest the demands for either cognitive processing or attention are lower at \qf or \rj when compared to \qs. The difference between \qs and \rj is consistent with the results by \citet{gwizdka2010distribution} and \citet{shovon2015search}. 

The difference between \qf and \qs further distinguishes these two EOIs. The high TAR and RPD at \qs are potentially due to the simultaneous cognitive activities for recalling information, and forming and expressing the query. If the task involves typing, it may also require additional effort to coordinate hand movements. However, a large deviation for \qf in Figure~\ref{fig:cogn_results} might indicate disengagement. Some participants disclosed that they do not always think about the query during given seconds. This was also a potential issue reported in \citet{moshfeghi2018search}'s experiment.

\paragraph{\hcog(4): \rj versus \ifn} No significant difference is found for TAR ($W_{TAR}=135$), but RPD is significantly lower at \ifn ($W_{RPD}=15, p<.001$) when compared to \rj. 
The TAR result may arise because both EOIs involve language comprehension, memory retrieval, and appraisal, and the cognitive demands for these tasks are not substantially different. 
On the other hand, even though the cognitive processes involved in both scenarios are similar, participants may exhibit more interest at \rj where they get search results that complete their knowledge gap. This heightened interest leads to greater engagement, and consequently, more directing of attention resources. As a result, RPD at \rj is higher than at \ifn. We further discuss this finding with the results of \haro(4) in the following section. As the causation might not primarily be attributed to cognitive load, \hcog(4) is partially supported.

\subsection{Arousal (\haro)}

\paragraph{\haro(1): \ifn versus \qf}
No significant difference is observed in SCL or HRV, while BAR is significantly different ($W_{BAR}=52, p<.05$). A higher BAR at \ifn compared to \qf may imply that the participants are becoming aware of their knowledge gap \cite{Dominika2022information} and consequently more alert in addressing it. 
The lack of significant difference in SCL or HRV suggests that alertness might have been mostly subconscious and not strong enough to reach a level of affective arousal. \haro(1) is partially supported. 

\paragraph{\haro(2): \qf versus \qs}
As shown in Figure~\ref{subfig:SCL} and \ref{subfig:HRV}, HRV is significantly lower at \qf as compared to \qs ($W_{HRV}=1, p<.001$). SCL is lower at \qs than at \qf, although the difference is not significant.
When a person experiences a high level of arousal, SCL will increase while HRV will decrease \cite{mohammadpoor2023arousal, pham2021heart}. These findings suggest arousal is relatively higher at \qf than \qs. Thus, \haro(2) is rejected.

\paragraph{\haro(3): \qs versus \rj}
Significant differences are observed for both SCL ($W_{SCL}=58, p<.05$) and HRV ($W_{HRV}=3, p<.001$). In addition to the lower SCL and higher HRV at \qs than \rj, as demonstrated in Figure~\ref{subfig:SCL} and \ref{subfig:HRV}, indicating arousal is lower at \qs than \rj, thereby supporting \haro(3).

\paragraph{\haro(4): \rj versus \ifn}
HRV is significantly longer at \rj than \ifn ($W_{HRV}=34, p<.01$), suggesting a higher level of arousal at \rj. This finding can be linked with the observation for \hcog(4) in Section~\ref{subsec:result_cog}, where a significantly larger RPD at \rj than \ifn. This elevated RPD is likely due to enhanced interest and engagement, which in turn translates to heightened arousal. Therefore, the result in HRV further rejects \hcog(4) and supports \haro(4).

\subsection{Valence (\hval)}

Our conclusions around valence mostly rely on FAA. 
The statistical results of FAA fail to reject the null hypothesis of any pair. This suggests valence between EOIs is not substantially different; \hval~s are rejected.
These results may add up to a balance of conflicting feelings, e.g. expectation and anxiety \cite{savolainen2015interplay}.
Refer to Figure~\ref{subfig:FAA}, all EOIs have FAA averages and medians nearly 0, implying neutral valence levels. A closer look at its components (Figure~\ref{subfig:FAA_components}) reveals that the right alpha power is relatively higher than the left at \ifn. Higher right alpha usually represents withdrawal motivation, associated with negative valence \cite{ramirez2012detecting, matlovivc2016emotion}. Right alpha is still slightly higher than left alpha at \qf. At \qs and \rj, the alpha levels are balanced. It might suggest valence tends to shift from negative to positive when participants figure out search queries and find relevant information.

\section{Discussion \& Limitations}
This study aims to characterize the information seeking process from 3 aspects, in relation to 3 physiological constructs: \emph{cognitive load}, \emph{affective arousal}, and \emph{valence}. Physiological signals are captured to compute the indexes infer these constructs. Our hypotheses are primarily built upon the findings of prior work in neurophysiology 
\cite{moshfeghi2018search} and behavioral analysis \cite{gwizdka2010distribution} (see Section~\ref{sec:hypotheses}).

Measured through TAR and RPD, \textit{cognitive load} shows statistically significant differences across search stages, supporting hypotheses \hcog(1--3). A noteworthy reversal between TAR and RPD at \ifn and \qf offers insights into \citeauthor{moshfeghi2018search}'s findings \cite{moshfeghi2018search}. At \ifn, goal-directed brain functions are more activated, aligning with the elevated TAR, while attention-directing functions are more activated at \qf, reflected in the higher RPD. Then, our results further distinguish between \qf and \qs, that \qs is more demanding. However, this is likely to be influenced by disengagement caused at the 10-seconds \qf. Moreover, \qs requires higher cognitive loads than \rj -- observed in both TAR and RPD -- which aligns with \citeauthor{gwizdka2010distribution}'s findings \cite{gwizdka2010distribution}.
Lastly, between \rj and \ifn, heightened RPD at \rj may be attributed to increased interest, along with engagement and arousal. This aligns with \citeauthor{paisalnan2021towards}'s findings \cite{paisalnan2021towards} of similar but more demanding cognitive processes at \rj compared to \ifn.

Three physiological indexes, BAR, SCL, and HRV, are used to measure \textit{affective arousal}. The hypotheses \haro are primarily validated with SCL and HRV. For \textit{affective valence}, the results from FAA fail to validate any hypotheses \hval, with no significant differences observed. But some variations are seen when examining the components of FAA, i.e., left and right frontal alpha. 
At the beginning of a search (\ifn), the elevated TAR and BAR suggest a knowledge gap is updated to the awareness. Additionally, right frontal alpha dominance implies a relatively negative valence. These physiological signs might infer that the \textit{feeling of uncertainty} primarily stays in a cognitive state, without the corresponding emotional reactions. Thus, emotional responses as a sign of need-to-search might not be effective. 
Arousal decreases and valence tends to be neutral at \qf, suggesting the participants are planning the action. 
Then, at \qs, arousal is lower compared to \qf.
This suggests that the pleasantness of being able to act, the expectation of success, and increased confidence in finding relevant results. Or it might because the participants were disengaged at \qf then back to the task at \qs. 
However, with no significant difference found in FAA, we cannot infer whether the feeling is positive or negative.
Consequently, when relevant search results are presented at \rj, arousal further increases than \qs. These findings are related to the reward-seeking feelings discussed in previous work \cite{moshfeghi2018search,lopatovska2014toward}.

Although \ifn and \rj may share similarities, our findings of significant differences in HRV and RPD, and observations of higher FAA, further highlight the difference between these two stages. The affective difference could be attributed to different appraising criteria \cite{nahl2007social}, such as having sufficient knowledge to solve the problem or accepting the found answers. Yet, the result of FAA is insufficient to conclude a positive emotion at \rj, so levels of satisfaction cannot be inferred from FAA. The results might relate to interest and curiosity.
The difference in RPD can be transformed as the web-logging behaviors, such as longer dwell time on the relevant results.

Like all experimental studies, our investigation is subject to limitations. Firstly, although we attempted to eliminate confounding variables, other factors, such as users' search skills and prior domain knowledge, may have influenced the results. 
Secondly, the indexes derived from physiological signals are insufficient to disclose all of the raw information they capture. The intricacies between cognitive load and emotions make it difficult to entirely separate their effects in physiological signals. In the next phase of this work, we plan to explore pattern analysis with machine learning models \cite{ji2023examining}, incorporating all features extracted from the signals. 
Despite this limitation, the specific indexes we used are based on empirical findings in the literature, which can provide a more robust description of subconscious user behavior. 
We note that the temporal relationship between search stages may influence physiological signals. However, for the sake of brevity, this paper compresses the entire signal into a single value, omitting the temporal relationship. Alternative aggregation methods will be explored, such as dividing into three equal-length segments -- beginning, middle, and end -- as applied by \citet{Gwizdka2017temporal}, or performing sophisticated temporal analysis of the entire signal, as shown by \citet{van2019analyzing}.

\section{Conclusion}
This study aimed to characterize user behaviors during an information seeking process using physiological signals. Our experiment focused on a scenario of searching for unknown knowledge and understanding a topic, i.e., searching to fill a knowledge gap. A lab user study was conducted to collect physiological signals through four search stages involving: the realization of Information Need (\ifn), Query Formulation (\qf), Query Submission (\qs), and Relevance Judgment (\rj). The \emph{cognitive load}, \emph{affective arousal} and \emph{valence} were analyzed using well-established indexes derived from the signals. 

Our results indicate that cognitive demands are higher, but attentional resources are lower at \ifn compared to \qf.
At \ifn, a slight rise in alertness might capture the recognition of the knowledge gap. But this response does not elicit any negative affective feelings, at least not to the extent that peripheral signals were able to detect in our experiment. 
Next, cognitive load is more intense at \qs than at the previous (\qf) or subsequent stage (\rj), which supplements the findings by \citet{gwizdka2010distribution} and \citet{shovon2015search}. This further indicates that simultaneous cognitive processes are highly demanding at \qs, potentially explaining higher affective feelings than at \qf. Finally, our results indicate that affective feelings are more active at \rj. Compared to \ifn, the incremental feelings and attentions at \rj suggest greater interest, engagement, and curiosity as the results resolve the searcher's knowledge gap.

This study extends the existing understanding of how users engage in information seeking processes by complementing existing theories and observational studies with the characterization of search stages using physiological signals.
Our findings serve as a baseline for future experiments investigating affective and cognitive feedback, as well as physiological signals, for search interactions. There is a growing interest in employing wearable physiological sensors in search systems due to their mobility, decreasing cost, and information-rich advantages \cite{schen2023informatics}.
By better understanding the intrinsic states of searchers in a continuous process, 
our proposed methodology can contribute to improving the overall search experience and devising real-time solutions. 
We believe our experimental setting -- validated in the context of known information seeking models -- can help characterize cognitive load and affective arousal in less established IIR settings, including Large Language Model-based conversational search.

\begin{acks}
This research is supported by the \grantsponsor{ARC}{Australian Research Council}{https://www.arc.gov.au/} (\grantnum{ARC}{DE200100064}, \grantnum{ARC}{CE200100005}).
\end{acks}

\newpage

\balance
\bibliographystyle{ACM-Reference-Format}
\bibliography{reference}

\end{document}

%% file: macros.tex
\newcommand{\cmark}{\ding{51}}%
\newcommand{\xmark}{\ding{55}}%
\newcommand{\myparagraph}[1]{\vspace{0.5\baselineskip}\noindent{\textbf{#1}}.~}

\newcommand{\task}[1]{{\small\sffamily #1}\xspace}
\newcommand{\eyeopen}{\task{EO}}
\newcommand{\eyeclose}{\task{EC}}
\newcommand{\ifn}{\task{IN}}
\newcommand{\qf}{\task{QF}}
\newcommand{\qs}{\task{QS}}
\newcommand{\rj}{\task{RJ}}

\newcommand{\ifnmath}{IN}
\newcommand{\qfmath}{QF}
\newcommand{\qsmath}{QS}
\newcommand{\rjmath}{RJ}

\newcommand{\hvalmath}{H_{val}}
\newcommand{\haromath}{H_{aro}}
\newcommand{\hcogmath}{H_{cog}}

\newcommand{\hval}{$\hvalmath$}
\newcommand{\haro}{$\haromath$}
\newcommand{\hcog}{$\hcogmath$}